\documentclass[aps,prl,twocolumn,groupedaddress]{revtex4}
\usepackage{graphicx}

\begin{document}

\title{  A physical basis for the phase in Feynman path integration}
\author{G. N. Ord}
\email[ corresponding author ]{gord@acs.ryerson.ca}
\affiliation{Department of Mathematics \\
Ryerson University\footnote{\jobname}\\
Toronto Ont.}
\author{J. A. Gualtieri}
\email{gualt@backserv.gsfc.nasa.gov}
\affiliation{Applied Information Sciences Branch\\
Global Science and Technology\\
Code 935 NASA/Goddard Space Flight Center\\
Greenbelt MD.}
\author{R. B. Mann}
\affiliation{Department of Physics \\
University of Waterloo\\
Waterloo, Ont.}
\date{\today}

\begin{abstract}
In the path integral formulation of quantum mechanics, the phase factor $%
e^{iS(x[t])}$ is associated with every path $x[t]$. Summing this factor over
all paths yields Feynman's propagator as a sum-over-paths. In the original
formulation, the complex phase was a mathematical device invoked to extract
 wave behaviour in a particle framework. In
this paper we show that the phase itself can have a physical origin in time
reversal, and that the propagator can be drawn by a single deterministic
path.
\end{abstract}

\maketitle







Quantum mechanics is an empirical theory with rules that have evolved in
response to increasingly precise experimental results. Although the
observational accuracy of quantum mechanics and its derivatives is beyond
question, the theory itself lacks a physical basis. Unlike classical
mechanics or relativity theory, quantum mechanics is not a mathematical
elaboration of any known set of physical principles. Indeed, whether or not
the theory describes an observer-independent physical reality is an open
question, with strongly divergent views within the physics community.

The closest that conventional quantum mechanics comes to a physical
`picture' of an underlying dynamical process is through the path-integral
formulation. In Feynman's space-time approach\cite{FeyHibbs}, a phase
angle $S(x[t])$ is associated with every continuous path $x[t]$, where
$S$ is the classical action of the path in units of $\hbar$. Each path is
then given the statistical 'weight' $e^{iS(x[t])}$ and the sum over all
such weights yields Feynman's propagator for the system.

Feynman's formulation is justly famous for its utility and intuitive appeal.
However the formulation falls short of providing a `realistic' basis for QM
on at least two counts. First of all the physical origin of Feynman's phase
is unknown. The complex algebra induced by the phase factor $e^{iS(x[t])}$
is responsible for quantum interference, but exactly \emph{how} a physical
path could contribute such a phase is not specified. Secondly, the
sum-over-paths is a sum-over-histories, each history being important to a
single  physical particle. In the classical analog, the Wiener integral, we
can imagine replacing the sum-over-histories by an ensemble average with the
usual appeal to classical concepts of probability. A classical particle
needs only the local information of its own Wiener path to obey, on average,
the diffusion equation. This is not the case for a quantum particle, which
needs non-local information in order to obey, on average, the Schr\"{o}%
dinger equation.

In this paper we address issues of the physical origin of phase and the
many-to-one aspect of Feynman paths. We propose a simple scheme where a
single particle `draws' a propagator on space-time much like an electron
beam draws an image on a cathode ray tube. In the case of the propagator,
the horizontal scan corresponds to a particle moving forward in $t$, the
horizontal retrace corresponds to the particle moving backwards in $t$.
In the propagator case the retrace `beam' is active and cancels any pixels
at intersections with previous forward sweeps. The forward and backward
sweeps oscillate about each other in a precise way at the Compton
frequency. It  is this oscillation that results in the appropriate algebra
of complex numbers. The central role of `zitterbewegung' in the model is
similar in character to that proposed by Hestenes
\cite{Hestenes83,Hestenes03}.

  The conceptual expense of this scheme is that
we have to allow the particle the capability to move in both directions in
time, the reversed time part being essential for quantum interference.
However, allowing this, the result is a simple, deterministic, ontological
model for a one dimensional propagator that provides a transparent
\emph{mechanism} for wave-particle duality.

We start with the Feynman Chessboard model\cite{
FeyHibbs,Gersch81,JacSchulman84,GJacobson84,gord92chess,noyes}. This was
Feynman's path-integral prescription for a propagator for the Dirac
equation. In units with $\hbar=c=1$ the kernel for propagation from $a$ to $b
$ is:

\begin{eqnarray}
K(b,a) &=&\sum_{R}N(R)(i\epsilon m)^{R}  \label{foursum} \\
&=&\left( \sum_{R=0,4,\ldots }N(R)(\epsilon m)^{R}-\sum_{R=2,6,\ldots
}N(R)(\epsilon m)^{R}\right)  \nonumber \\
&&+i\left( \sum_{R=1,5\ldots }N(R)(\epsilon m)^{R}-\sum_{R=3,7,\ldots
}N(R)(\epsilon m)^{R}\right)  \nonumber \\
&=&\Phi _{+}+i\,\Phi _{-}.  \nonumber
\end{eqnarray}%
where the sum is over all $R$-cornered paths with step size $\epsilon $ and $%
N(R)$ is the number of $R$-cornered paths.  In Eqn.(\ref{foursum}) we see
that $i=\sqrt{-1}$ serves two purposes. It distinguishes right from left
through the partition of the sums into real and imaginary components $\Phi
_{\pm }$. It also constructs interference effects through the periodic
subtractions within each component $\Phi $. As has been pointed out in
previous work, \cite{gord01a,gnorbm03a,gnorbm03b} these subtractions can be
given a realistic basis by covering the chessboard ensemble of paths by a
single `Entwined Path' (EP) that traverses the subtracted portions of the
chessboard paths in the $-t$ direction, thereby giving the interference
effects a physical origin.

\begin{figure}[tbp]
\includegraphics[scale = .8]{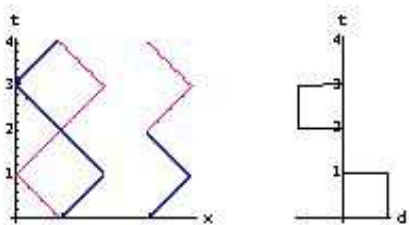}
\caption{An entwined path with the right envelope/chessboard path. The
left-most graph is a short 'fibre' in which the direction of traversal is
colour coded, blue (thick line) for forward in $t$, red (thin line) for
backwards in
$t$. The right envelope is the right half of the fibre. The graph on the
right represents a density $(d)$ of right-moving (adolescent) particles
from the chessboard path. The density is 1 for the forward-$t$ portion
and -1 for the reversed-time portion. The gap between the two arises
because there is no right-moving particle in the right envelope from
$t=1$ to
$t=2$. }
\end{figure}

\begin{figure}[tbp]
\includegraphics[scale = .55]{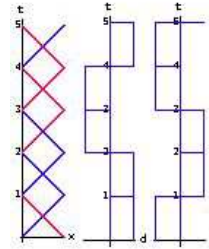}
\par
\caption{On the left are two right envelopes from a concatenated pair of
entwined paths, where one envelope is displaced by 1/4 cycle (one time
step) from the other. They form a density with a step function form. 
In the center is the density resulting from the adolescent particles
Eqn.(\ref{stepa}) and on the right is the density for senescent
particles.  }
\label{hists}
\end{figure}

For our purposes the above EP scheme has two limitations that we seek to
modify. First of all the scheme is stochastic. Since any stochastic process
can be mimicked by a deterministic process that preserves statistical
averages, we seek to do this in a simple way. The stochastic version of EPs
converges slowly to the appropriate propagator because each space-time step
involves a stochastic decision. This makes simulations slow and
computationally expensive. Both of these limitations can be removed by a
judicious choice of an underlying dynamical process that we now describe.

To facilitate removal of the stochastic component we consider what we shall
call `velocity eigenpaths'. These are deterministic EP's where the 
geometry of each closed loop is identical within a given path. A $v=0$
eigenpath is sketched  with the associated right envelope
(righthand path) over a period of a single cycle in the left side of Fig.
1.. As in the stochastic case we need only consider the left or right
`envelope' for counting paths. Each envelope contributes alternately to
right-moving and left-moving particles, the contribution being $\pm 1$
depending on the direction in $t$ of the original entwined path. If we
look, for example, at the right-moving states, the single right envelope
creates a square-wave density with periodic gaps where the particle has
switched direction. Density is defined as the total number of paths
through each space-time point weighted by +1 for forward in time and -1
for backward in time. We count paths according to direction because each
small loop can be interpreted as a creation of a virtual pair followed by
an anihilation of the same pair at the end of the loop. We only use one
envelope for counting because, in one-dimension, the other envelope
contains the same information. We call right-moving particles on the
right envelope adolescent (just post-creation) and left moving particles
senescent (pre-annihilation). The relationship between these two
populations has an interesting feature due to the geometry of the paths.
The adolescent density for the sketched path is:

\[
u_{A}(t)=\left\{ 
\begin{array}{ll}
1 & \mbox{if mod($t$,4) $ \leq 1$} \\ 
-1 & \mbox{if $2 < $ mod($t$,4) $ \leq 3$} \\ 
0 & \mbox{otherwise}%
\end{array}%
\right. 
\]%
where we have not specified boundaries for $t$. In Figure (1), $t$ starts at 
$t_{0}=0$ and ends after one cycle at $t_{R}=4$ but we could choose other
values for start and return times. Note that the senescent density for our
path is simply $u_{S}(t)=u_{A}(t-1)$. 
\begin{figure}[tbp]
\includegraphics[scale = .55]{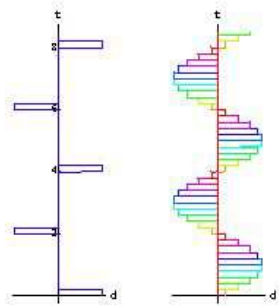}
\par
\caption{The periodic delta function of Eqn.(\ref{eqn4}) for a single
quartet of EP's (cord) on the left. Sequential translations of cords are
used to construct an approximation to $\sin (\protect\pi x/4)$ in the right
hand figure. }
\label{histsb}
\end{figure}

To make contact with Feynman's phase, we would like to make the densities
drawn by our entwined path more obviously connected to $%
e^{-imt}$, since this is essentially the carrier wave generated in the
Chessboard model. To do this deterministically we shall consider a three
stage process in the construction of a suitable entwined path. We first of
all construct a `fiber', a single entwined pair as illustrated in Fig. 1. We
then concatenate four fibers to produce a `cord'. A cord produces a periodic
sequence of delta functions on space-time, as in the left half of Fig.(3).
Finally we concatenate cords to build a `cable' in such a way that the
resulting space-time densities approximate $e^{imt}$. At all steps in this
process we add paths by concatenation \ldots the resulting paths from fiber
to cable form a single continuous space-time path.

Proceeding, notice that if we concatenate two fibers shifted by one quarter
period ( Fig.(2) ) the resulting space-time densities for the two
right-envelopes are  square waves. For the adolescent density we have: 
\begin{equation}
W_{a}(t)=u_{a}(t)+u_{a}(t-1)  \label{stepa}
\end{equation}
The density of senescent particles is then just $W_{a}(t-1)$. If we choose a
lattice spacing $\epsilon =4/(2n)$ we can convert the square-wave densities
to a periodic array of lattice delta-functions by concatenating two more
fibers shifted one lattice spacing away from a half-period shift of the
original two. 
This will give an adolescent density function: 
\begin{equation}
\delta _{a}(t)=W_{a}(t)+W_{a}(t+2-\epsilon )  \label{eqn4}
\end{equation}%
This is pictured on the left in Fig.(3). It is a periodic chain of discrete
delta functions with alternating signs. The alternating signs allow us to
simply concatenate paths, corresponding to a physical process of \emph{adding%
} real paths, while constructing a real density in which \emph{subtraction}
plays an important role. Adding paths in the Wiener context does not allow
for this, and cannot accomodate self-interference as a result. The same
concatenation that produces $\delta _{a}(t)$ for adolescent particles
automatically produces $\delta _{s}(t)=\delta _{a}(t-1)$ for senescent
particles.

Now we repeat the concatenation process shifting the temporal origin  by $%
\epsilon $ each time. We do this $n$ times and at the $k$'th shift we
concatenate $[|M\sin (\pi k/n)|]$ cords. Here M is a large integer chosen to
extract a specified number of decimal places accuracy from the trigonometric
function. The resulting density is as sketched, for $n=10$, $M=20$ in
Fig.(3) on the right. The scenescent density is identical in form but lags
in phase by $\pi /2$.

It is clear from this procedure that we can get as close as we like to a
representation of the trigonometric functions over any fixed number of
cycles, simply by carefully choosing the EP concatenations. It should also
be clear that there are many ways of assembling a cable to draw the
two components of $e^{imt}$ on space-time. We have chosen a procedure that
is transparent in its ultimate outcome, but unlikely to occur
 in the real world. However, the point here is not the
artificiality of the construction method, it is the fact that it is a
method that
\emph{could}  be used by a single point particle with no non-classical
properties to 
\emph{encode} quantum information in space-time. This is in contrast to the
conventional formulation of Feynman which is explicitly a technique whereby
we use phase to
\emph{extract} quantum information from Wiener paths. Unless we take a
many-worlds interpretation of quantum mechanics, those Wiener paths have no
direct physical counterpart.


To see how this scheme relates to the path-integral we recall that in the
non-relativistic limit,( $x/t<<1$, $t>>1/m$ ) the un-normalized propagator
from the Chessboard model may be written: 
\begin{equation}
K(x,t)=\exp (-imt(1-\frac{x^{2}}{2t^{2}}))
\end{equation}%
up to a constant phase factor. The non-relativistic propagator is simply a
low frequency signal sitting on top of the high-frequency 
zitterbewegung generated at the Compton frequency
$m$. If we write this
in terms of
$v=x/t$ this is 
\begin{equation}
K(x,t)=\exp (-imt)\exp (i\frac{mv^{2}}{2}t)
\end{equation}%
and we see that along a ray of constant $v$ we have a simple complex
exponential with a frequency reduced by a term proportional to the classical
action along the path. We can thus construct, along this ray, an EP
approximation to the complex phase factor simply by `writing' along the ray
with a suitable EP.

Thus, each point on our lattice space-time will lie on a single ray from the
origin, and will be written to by a unique set of EP's that will approximate
the appropriate phase factor. We simply have to sweep the EP's with
appropriate frequencies over all rays of fixed $v$. The propagator (4)
in principle allows one to construct any solution to Schr\"{o}dinger's
equation that is accessible to the path integral. A simple example is shown
in Figure(4). The apparent standing wave in the figure is the first excited
state of a particle in a periodic box. It was produced without
waves or quantization. It is one component of the density generated by 
two concatenated 
 velocity eigen-paths with the speed fixed at the appropriate square root of
the energy. In general, eigen-paths chosen to correspond to energy
eigenstates yield standing waves, while other speeds do not. Figure (4)
shows  explicitly a manifestation of the kind of wave-particle duality
available to velocity eigenpaths. The density registered by spacetime is
clearly a standing wave, yet the wave itself was constructed by a single
'particle' with a one-dimensional parameterization.
\begin{figure}[tbp]
\includegraphics[scale = .40]{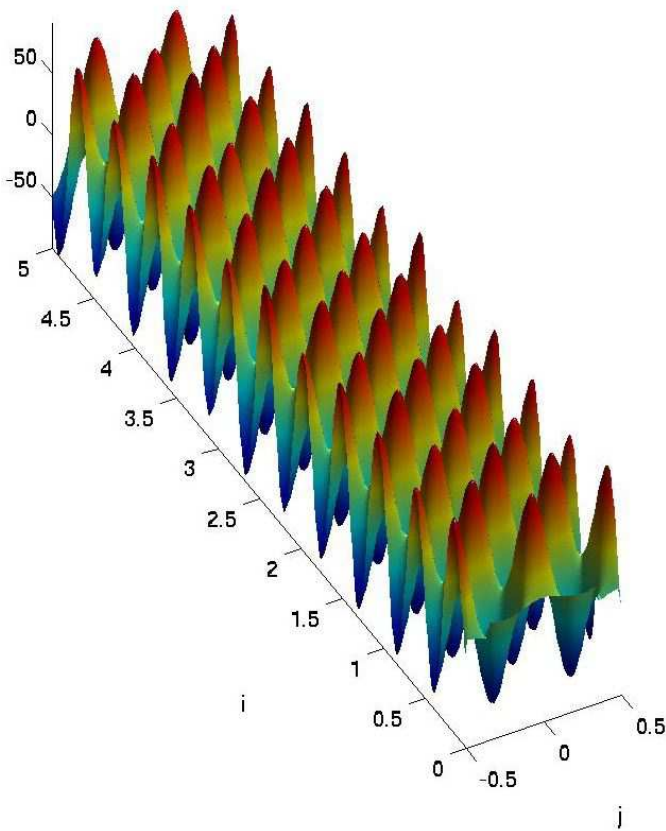}
\par
\caption{Entwined path right-moving density for a particle on a ring. The
axis labeled j is the $x$-axis, the axis labeled i is the $t$-axis. The
remaining axis records the density accumulation from a single entwined
path. The standing wave in the density is a result of the underlying
dynamical process (velocity eigen-path) and the initial condition at
$t=0$. }
\label{histsb}
\end{figure}

It is worth emphasizing the context of the above demonstration in light of
current views on quantum mechanics. All current versions of quantum
mechanics rely explicitly on some form of canonical quantization, usually
the explicit replacement of classical dynamical variables by operators. In
the path integral formulation, this process is replaced by the invocation of
phase. Both procedures work `for all practical purposes'\cite{JBell90a} but
the end result is  a \emph{recipe} for calculation, rather than a
description of an underlying particle motion.


The above demonstration suggests that the pursuit of a sub-quantum dynamic
giving  rise to wave-particle duality, even
when interactions are present, may  be a feasible task. This would call into
question the postulate that wavefunctions provide complete information
about the state of a physical system, if not operationally, then at least
conceptually.
The free particle propagator obtained above, which underlies all of
non-relativistic quantum mechanics, is easy to construct in any space-time
region using only a single entwined space-time trajectory, itself a simple
concatenation of cables. The only conceptual price we have paid for this
realist construction is that a particle's world-line can, in a restricted
way,  run backwards in time. The restriction, entwinement about the most
`recent' forward path, ensures the `causal' evolution of the conventional
propagator. The extra degree of freedom obtained by allowing both
directions in $t$ allows the particle to establish the non-local
information needed for wave-particle duality without the necessity of
intelligent particles, intelligent observers, or multiple universes.



\begin{acknowledgments}
This work was funded in part by the Natural Sciences \& Engineering Research
Council of Canada (GNO, RBM).
\end{acknowledgments}

\end{document}